\documentclass[a4paper,11pt]{article}
\pdfoutput=1 

\usepackage{jinstpub} 
\usepackage{indentfirst}
\usepackage{graphicx}

\title{On Proportional Scintillation in Very Large LXe Detectors}

\author[1]{P. Juyal,\note{Corresponding author.}}
\author{K. L. Giboni,}
\author{X. Ji}
\author{and J. Liu}

\affiliation{School of Physics and Astronomy, Shanghai Jiao Tong University,\\
Shanghai Key Laboratory for Particle Physics and Cosmology(INPAC), 800 Dongchuan Road, Shanghai, 200240 P.R. China\\}

\emailAdd{pratibhajuyal@sjtu.edu.cn}

\abstract
{The charge read out of a LXe detector via Proportional Scintillation in the liquid phase was first realized by the Waseda group 40 years ago, but at that time the technical challenges were overwhelming. Although the tests were successful, this method was finally discarded and eventually nearly forgotten. For present day large LXe Dark Matter detectors, this approach was not even considered. Instead the Dual Phase technology was selected despite many limitations and challenges. In two independent studies the groups from Columbia University and Shanghai Jiao Tong University reevaluated Proportional Scintillation in the liquid phase. Both established the merits for very large LXe detectors, but the Columbia group also encountered apparent limitations, namely the shadowing of the light by the anode wires and a dependence of the pulse shape on the drift path of the electrons in the anode region. The discrepancies between the two studies, however, are not intrinsic to the technique, but a direct consequence of the chosen geometry. Taking the geometrical differences into account the results match without ambiguity. They also agree with the original results from the Waseda group. With the technical problems solved, the path is now open to use this method in future large LXe TPCs.}


\begin{document}
\maketitle
\flushbottom

\section{Introduction}
The liquid xenon (LXe) Time Projection Chamber (TPC) at present seems to be the detector of choice for the search of Dark Matter (DM) in the form of Weakly Interacting Massive Particles (WIMP) at high masses (around 50 GeV/c$^2$ and above). It is a powerful instrument, but DM search still challenges it to its limits. The energies to be detected are very small (< 50 keV$_{ee}$), i.e. all events are point-like and any structure of the interaction is below the spatial resolution. It is only due to the excellent qualities of xenon that one can use it even in this range. But the energy is so small that a conventional charge read out with a Charge Sensitive Amplifier (CSA) is impossible. For these cases, Dolgoshein ~\cite{A} developed the "Dual Phase" method. The ionization electrons are extracted from the liquid and generate proportional scintillation in the gas above. The weak light signal can be detected via the noiseless amplification in a Photo Multiplier Tube (PMT).

Of course LXe is also known as an efficient scintillator. The light at such low energies can be detected using PMTs or other VUV-photon sensors. But, the photons are emitted isotropically. An efficient 4$\pi$ photo sensor array around all the volume is challenging and expensive for a ton scale detector. To alleviate this problem the side walls are usually made reflective with Teflon ~\cite{B} to detect as many photons as possible in two PMT arrays, one above and one below the active volume. Of course it is also possible to use the light alone to detect WIMPs like in the XMASS experiment ~\cite{C} but the simultaneous measurement of  the charge provides important information which can be used e.g. to improve the energy resolution.

Higher efficiencies and better resolution can be achieved when more information is available about an event, e.g. by detecting both, the light and the charge. Naturally the above mentioned Dual Phase detector can measure the charge with proportional scintillation, but also detects the direct scintillation light. Both quantities are a measure of the amount of energy deposited by the event. One advantage in having both measurements is an improved energy resolution of the detector. This takes advantage of a known anti-correlation ~\cite{D} between charge and light. Considering the anti-correlation improves the energy resolution of the detector far beyond what would be expected from the two measurements separately.

There is, however, an additional benefit beyond energy resolution, and this is background reduction. WIMPs interact with the entire Xe nucleus giving it a recoil which in turn produces the ionization signal. This class of events we call nuclear recoil (NR). Most of the background events are caused by X- and $\gamma$-rays with an electron recoiling, called electron recoil (ER). The two event types can be distinguished by the ratio S2/S1 of the charge signal (S2) and the light (S1). A cut at an appropriate value of S2/S1 suppresses the $\gamma$-ray background by a factor of nearly 100.  Thus the dual phase principle is an elegant way to measure the deposited charge, but it also introduces severe restrictions on the geometry and the operating point of the detector. Also, there are requirements for the mechanical precision of the detector elements. These conditions are easy to fulfill in a small detector, but it becomes very challenging when the diameter of the anode grows beyond 1 m. A review of the properties of detectors are described in ~\cite{E} and ~\cite{F}

Proportional Scintillation in liquid xenon is known. This was established ~\cite{G}~\cite{H} by the Waseda group in 1979. The field strength must be of order 400 kV/cm to reach the proportional regime. Such fields occur in the 1/r field around thin wires. Although the Waseda tests were a great success, the group encountered too many technological difficulties. Furthermore, the tests were all performed at higher energies where Charge Sensitive Amplifiers were available. With so many difficulties and no real advantage, the results of these tests were practically forgotten. In the mean time the technological challenges have been solved, and we are free to use proportional scintillation in a single phase detector with many benefits for the overall performance. 

Recently Proportional Scintillation in the liquid was again evaluated in two independent studies, one at the Columbia Astrophysics Lab (CAL) ~\cite{I} and the other at Shanghai Jiao Tong University (SJTU) ~\cite{J}. Both studies confirmed the Waseda results, but the CAL results seemed to imply some severe drawbacks, incompatible with high-resolution measurements. In a critical comparison we can trace back the origin of the problems to an unfavorable geometry of the test detector. The discrepancies can be fully explained, and the problems can be avoided in the design of a DM detector. In the future, a single phase design measuring the proportional light in the liquid could replace the Dual Phase design. Thus we can develop detectors with less compromises, easier to design and operate, but still with better performance. This might turn out to be a crucial advantage in the design of future, even larger DM detectors ~\cite{K}.

\section{Dual and Single Phase Detectors}
All LXe Dark Matter detectors presently in operation are dual phase with the exception of XMASS, which is only a scintillation detector. Concerning the nomenclature, in the literature the term single phase often assumes a scintillation only detector. In the following, however, we use single phase describing any LXe detector not using the gaseous form of xenon. Of course, we are mostly interested in TPC detectors which measure both, charge and light. Thus, in the following we deal with single phase detectors which measure the charge with proportional scintillation in the liquid.

\subsection{Dual Phase Method}
\begin{figure}[!hbp]
	\centering
	\includegraphics[width=1\textwidth]{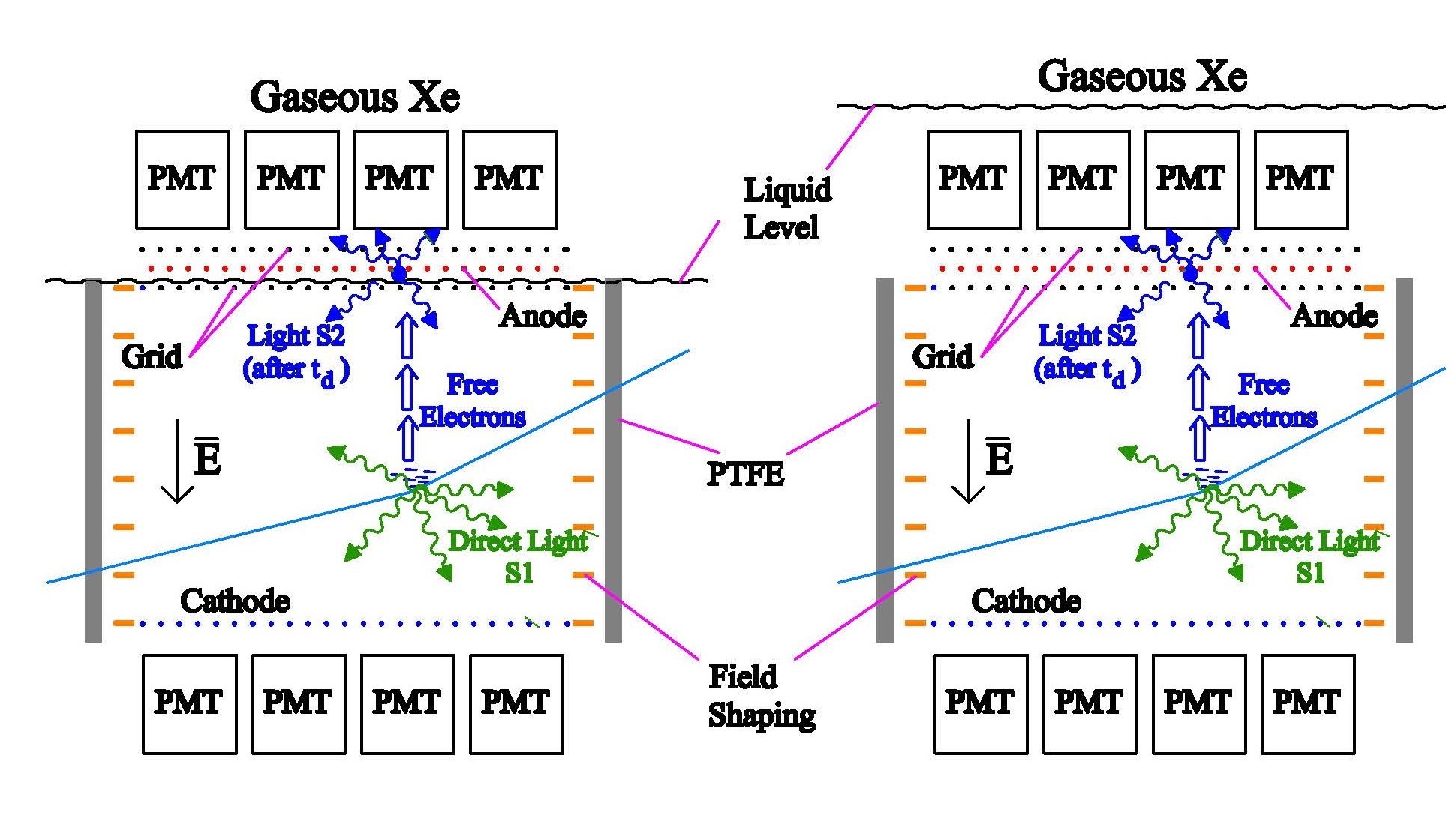}\\
	\caption{Schematic working principle of a dual phase (left) and a single phase (right)	LXe detector. The differences are mainly the position of the liquid level and the diameter of the anode wires.}
	\label{fig 1}
\end{figure}

The Dual Phase principle is often explained with a drawing similar to Fig~\ref{fig 1}. The left part of the figure shows the active liquid xenon volume with a sample interactions producing scintillation photons and free drifting electrons. The direct light (S1) is then detected by the two PMT arrays below the cathode and above the anode in the gas phase. The charges drift in the applied electric field and are extracted from the liquid. In the strong homogeneous field above the liquid level they produce Proportional Scintillation (S2), which in turn is also seen by the two PMT arrays. The right hand side depicts the equivalent single phase detector to be discussed later. The main differences are the anode wire diameter (20 $\mu$m instead of 100 $\mu$m, and the liquid level which is now above the top PMT array.

The Dual Phase method is an elegant way around the lack of adequate charge sensitive amplifiers. The technique is so sensitive that even a single drifting electron can be detected ~\cite{L}. Practically, however, this simple principle also imposes stringent limitations on the detector design. For example, the anode potential controls two different processes, the extraction from the liquid and the proportional scintillation in the gas gap. The potential has to be sufficient to assure a good extraction efficiency, but high fields result in a very large gain and potentially the formation of electron avalanches or saturation of the read out electronics. The best choice for the anode potential will be a compromise. For these operating conditions the PMTs have to detect everything from the weak direct scintillation light at low energies to the intense proportional light at high energies. The dynamic range of the read out will eventually define the effective energy range of the measurements. 

Fortunately there is a combination of design parameters with an adequate performance. The gap between liquid level is typically kept at 3-5 mm, and the anode potential around 5 kV. Naturally, the anode must be transparent to the scintillation light, and either stretched wires or meshes are used. Of course, within the required precision the wires must not sag under the influence of gravity or the electric field. Meshes are considered simpler, but the manufacturing by either etching or electro-forming does not produce round cross sections. At the sharp corners of the conductors we have a 1/r field much higher than in the parallel gap. This might lead to local avalanche formation, or worse breakdowns. Localized avalanche formation in a fraction of the events reduces the energy resolution. Evidence can be observed in the pulse shapes of the charge signals.

The observed signals from the PMTs are an initial very fast pulse of the direct light at t = t$_{0}$. The subsequent pulses are from the proportional light. With the known drift velocity the arrival time relative to t$_{0}$ is a measure for the z-coordinate of the interaction. In an event there is one S1, but there can be more than one S2 pulses, e.g. when a $\gamma$-ray interacts via Compton scattering.

As mentioned earlier we observe the event types, NR for WIMP candidates and ER for $\gamma$-ray background. The light from both types has the same wavelength, but the decay time and the fraction of charge to light (S2/S1) production is different. The details of the signal formation are discussed in a recent study ~\cite{M}. Practically, the discriminator against ER events is the ratio S2/S1. The S1 pulse shape is determined by the original light pulse and pulse shaping by the cable connections. It is a very fast, prompt signal. The S2 pulses are produced during all the time the electrons cross the gap between liquid level and anode. Therefore, the S2 pulse shape should be a 1 $\mu$s wide flat top pulse. With the usual 10 nsec binning the statistical fluctuation in each bin are large, and it is difficult to detect any structure within the recorded pulse.

This very short description already points to the main drawbacks of dual phase. The active volume must include the liquid level, i.e. the S2 depends on the path length between liquid level and anode. Therefore, the anode must be completely parallel to the liquid level. Any deviation will make the S2 position dependent. Of course the gap between anode and liquid level must be controllable, i.e we must be able to adjust the liquid level. This is easy with small detectors, but when the diameter of the anode goes beyond 1 m and the mass of the detector is more than a few hundred kg, it becomes a challenge.

We also have to remember that the anode and its HV connection are in the gas phase. The anode wires or meshes must be entirely plane despite gravity and the forces from the electrical field of order 10 kV/cm. Any disturbance of the field can lead to spurious discharges. For long run periods the anode HV is kept below the optimal value to avoid breakdowns, but this means a reduced extraction efficiency.

Detecting the location of the S2 pulses with a large PMT array determines the position of the interaction with good precision even with the granularity of 3'' PMTs. Thus we get a good spatial resolution of order few mm. However, this only applies for single site events. At high energy $\gamma$-rays prefer to Compton scatter with a small scattering angle, i.e. with low energy deposition. Afterwards they still have nearly the same energy and can Compton scatter again. We can reject such an event as ER if we can separate the two interaction sites. The position resolution in the anode plane (x-y) is very good with the PMT array, but the `Double Hit' resolution is not. We still can separate the two locations in the z-coordinate, i.e. in case they are separate in this projection. This separation is easy to apply, except for small distances due to the square shape of the S2 pulses of 1 $\mu$s duration with its statistical fluctuations.

Despite these and other drawbacks Dual Phase still provided good results even in very large detectors.

\subsection{Single Phase Method}
\setlength{\parindent}{5ex}The construction of a Single Phase looks quite similar to a Dual Phase, but with the liquid level outside the active volume. The anode structure is again an array of three electrodes, two shielding grids with the anode in the center, but all are immersed in the liquid. The liquid level is even above the top PMT array to reduce HV breakdowns on the PMT bases. The anode is made of stretched wires, normally 20 $\mu$m gold-plated tungsten. The arrangement of the electrodes resembles a Multi Wire Drift Chamber with two shielding grids sandwiching the anode. The event generation, charge drift, and S1 detection are identical to the before described Dual Phase operation. 

When the electrons approach the shielding grid they see the stronger field  ~\cite{N} ~\cite{O} of the anode and are guided by the field lines around the wires of the shielding grid. The anode potential causes a strong 1/r field around the thin wires. Close to the wire when the field strength exceeds 412 kV/m the electrons produce proportional scintillation. Above 725 kV/cm electron multiplication would set in. Since avalanches introduce additional statistical fluctuation the anode potential should be chosen such that this value cannot be reached before the electrons hit the wire surface. Practically the proportional scintillation only occurs very close to the wire, normally less than the radius value above the surface. There are no space charge effects since no positive ions are produced. And, there are no avalanches since the S2 photons cannot liberate electrons.

The S2 gain is different since now the photons are no more produced over the long distance between liquid level and anode. And the S2 pulse will be less than 100 ns long.

\subsection{Major Differences of DP and SP}
Naturally, the most obvious difference of a Single Phase detector is the absence of the liquid level in the active volume. There are many more differences when looking at the details. Keeping a large LXe DM detector in mind the differences in detector design are compiled in Table 1. Most of the differences can be turned into benefits either for the design or the operation phase of the detector.

\begin{table}
	\caption{Major Differences of Single Phase Operation and their benefits }
	\label{tab:2} 
	
	\begin{tabular}{p{6cm}p{8cm}}
		\hline
		Liquid Level not in active volume 	& No level control mechanism  \\
		& No leveling of the detector required \\
		& Electron can drift in any direction \\
		& Multiple drift regions are possible \\
		& No total internal reflection on liquid level \\
		& Active shield in front of Top PMTs \\
		& Waves or ripples have no effect \\
		\hline
		
		Anode within liquid& No more `spurious anode HV trips'  \\
		& No extraction efficiency \\
		\hline
		Multiple Drift Regions& Much shorter max drift distance \\
		& Reduced cathode HV  \\
		& Cathode HV far from PMTs  \\
		& Less electron attachment to impurities \\
		\hline					
		Clean Short Pulses(<100ns)& Better Compton recognition\\
		\hline	
		
	\end{tabular}
\end{table}

\section{Test Results and Discrepancies}
\label{Test Results and Discrepancies}

There were three independent evaluations of proportional scintillation in LXe. The original experiment by the Waseda group proved the existence of the mechanism although any application was highly discouraged by technological challenges. Two more recent studies aimed at evaluating the method for use in future large DM experiments one from CAL and the other from SJTU. The technological challenges were in the mean time eliminated.

The study at CAL used a simple geometry of a single 10 $\mu$m wire centered between two planes of stretched wires viewed by two large PMTs one above and one below the active volume. The wire arrangement was staggered, i.e. within the x-y plane the anode wire was in between the locations of two adjacent grid wires. In the SJTU approach three stretched wire grid planes formed the anode structure. The gold-plated tungsten wires were thicker with 20 $\mu$m and 50 $\mu$m similar to the standard wires in MWPCs. The grids were aligned such that the wires were located one behind the other in the three planes. Two arrays of four PMTs each were used to detect the light above the anode and below the cathode.

The results from all three studies overlap and would be in favor of a single phase approach. However, the CAL study also observed effects which would be not desirable in a DM detector.

\subsection{Triple Pulse Component}
\label{Triple Pulse Component}
\begin{figure}[!ht]
	\centering
	\includegraphics[width=1\linewidth]{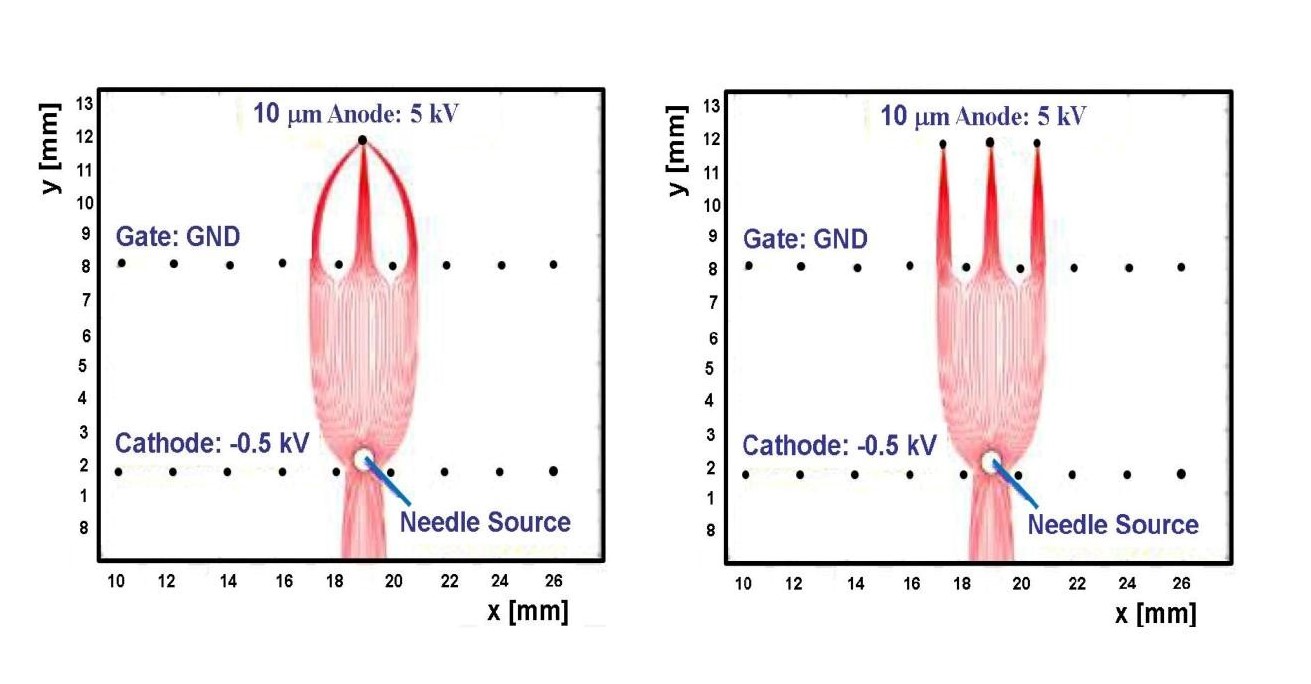}\\
	\caption{Drift paths of ionization electrons from a point source to the anode wire in the CAL experiment. (reproduced from ref. I with permission of the authors).  a. On the left, the original plot. i.e. the case of a single anode wire. The three different flight paths can be identified. b. On the right, the case with three anode wires. The plot was not redrawn, but modified by symmetry arguments. All the flight paths are now the same, but the two additional components are ending on the adjacent anode wires. }
	\label{fig 2}
\end{figure}

\setlength{\parindent}{1cm}
\indent
The CAL experiment observed the drifting electrons from a $^{241}$Am alpha source on the cathode. The range of the alphas is of order 20$\mu$m, i.e. the events are point-like. The ionization electrons drift towards the anode assembly following the field lines. Close to the shielding grid the field lines are focused to pass through the spaces in the grid towards the anode. Depending on the exact location of the alpha the drift path either goes straight up to the wire or passes through the adjacent spaces in the grid. The path is different in the three cases giving origin to different pulses. The different paths are shown Fig.~\ref{fig 2} on the left side. This plot is the original from the CAL publication.

However the implemented geometry is not a good approximation to a DM detector. There will never be only a single anode wire, but an array of wires like in an MWPC. Adding the next adjacent wires will change the field lines for the paths on either side. These electrons will be guided to adjacent anode wires. To evaluate the effect we do not have to recalculate the field lines but can deduce the changes by symmetry considerations. This is illustrated on the right in Fig.~\ref{fig 2}.b showing also the field lines if the adjacent anode wires would have been present. The drift path and thus the pulse shape will be indistinguishable from the central channel. This discrepancy in the results is therefore caused by using a single anode wire only.

\subsection{Shadowing}
\label{Shadowing}
With an anode wire diameter of 10 $\mu$m as in the CAL study  the drifting electrons will produce proportional scintillation once the 1/r field is above the threshold (412 kV/cm). Increasing the field strength would push the start of the region farther away from the wire, but as explained earlier the field strength at the surface must remain below the threshold for electron multiplication. Staying within these limits practically means that the maximum path length over which proportional scintillation is produced is very short. For easy estimations it is typically less than a wire radius. Naturally such a short light source at the surface of a wire will always cast a shadow.

\begin{figure}[!ht]
	\centering
	\includegraphics[width=0.8\textwidth]{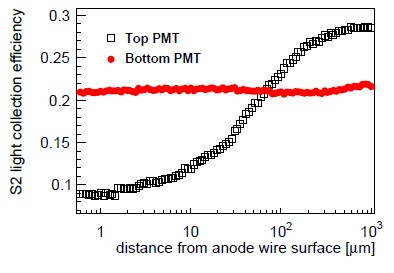}\\
	\caption{Light Observed on the bottom and top PMT from proportional scintillation in liquid xenon with a 10 $\mu$m wire in dependence of the distance from the wire. Note that the light is only produced very close to the wire(less than 5 $\mu$m away).}
	\label{fig 3}
\end{figure}

Fig.~\ref{fig 2} shows us that the electrons in the CAL geometry are hitting the wire from below. Obviously most of the light for the Top PMT array is blocked by the wire. The CAL group calculated the relative light on the Top and Bottom PMT in dependence of its distance from wire surface. This calculation is shown in Fig.~\ref{fig 3}, reproduced from the CAL publication. The Bottom PMT sees a constant amount of light, but the Top PMT not only has a much lower Light Collection Efficiency (LCE), but this even varies dramatically with distance. Although correct, this plot is misleading. The scale of the plot reaches out to 1 mm away from the wire. With their 10 $\mu$m wire, however, proportional light will be only produced in the first 5 $\mu$m. In this range the LCE of the Top PMT is low, but at least constant.

\begin{figure}[!ht]
	\centering
	\includegraphics[width=0.8\textwidth]{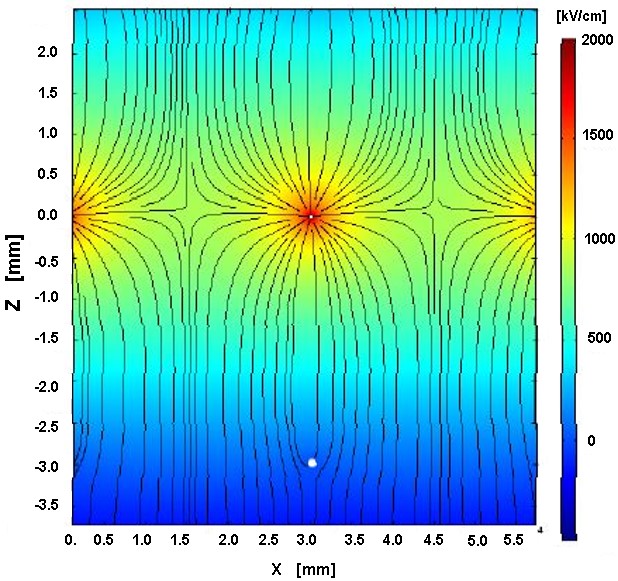}\\
	\caption{Field lines around the anode region for a single phase detector}
	\label{fig 4}
\end{figure}

Fig.~\ref{fig 4} shows the field distribution for the SJTU geometry with aligned wires. The electrons are deviated around the bottom grid wire and continue until they are bend towards the anode wire by the field lines from the top grid. There are no electrons hitting the anode from the bottom, i.e. in the region of maximum shadow for the Top PMT. Many of the electrons approach the wire from the side and a large fraction of the proportional light will be observed with the Top PMT. This is complemented with many more photons being seen by the Bottom PMT. Thus, both PMTs see an S2 pulse.

The S2 light is used for two different tasks, position reconstruction in the X - Y plane and total charge measurement. With the aligned arrangement of wires a sizable amount of photons are going to the Top array. Thus, the first task can be accomplished even with the Top PMTs alone. Additionally, earlier tests with PandaX I data showed that the S2 signal on the Bottom array alone is also sufficient to determine the event position. For the second task, the total energy measurement, we can add the Top and Bottom PMT signals. Thus, shadowing will not be a concern in a Single Phase detector, neither for the position determination nor for the total charge measurement.

\subsection{Expected Signal Size and S2 Gain}
Determining the operating conditions in a Dual Phase detector is not an easy task. The number of proportional scintillation photons depends on several parameters which cannot be chosen freely. Other considerations, like stability of operation and mechanical tolerances also enter in the optimization process. For example, the anode potential has to be of order 5 kV to efficiently extract the electrons into the gas phase, but such high voltages enhance the probability of spurious breakdowns at any imperfection of the structure including the connection in the gas. Thus the voltage is a compromise between extraction efficiency and stability of operation. The distance between anode and grid wires, typically 5 mm, is a further example. This gap is cut in half by the liquid level and the field above the liquid level is supposedly homogeneous. The gap cannot be made significantly smaller because of the mechanical tolerances in a large detector with more than 1 m diameter. And any deviation due to sagging wires or imperfections in leveling will have a large local effect on the S2 signal. Any gap width significantly larger would require an excessive anode voltage.

The situation in a Single Phase detector is very different. There is no extraction efficiency. The S2 production only depends on the first 5 - 10 $\mu$m around the anode wire. Field calculations show that displacing the anode wire by a full 1 mm does not significantly change the field in this region. Also, all HV connections are within the liquid. To first order we can optimize the anode voltage without compromising the performance of the detector.

Although it is convenient to have a strong S2 signal, all the light has to be observed by the same PMTs. The read out must be sensitive to the feeble S1 light at the lowest energies as well as the strong S2 light of the highest energies of interest. In a Dual Phase detector the S2 gain is of order 200 - 300 Ph/e$^{-}$. Such a high gain easily can cause the read out to saturate. The S1 light is produced deep inside the active volume, and a PMT will normally only see a few photons. Not so for the strong S2 light. It is produced at the edge of the active volume and many photons will hit the same PMT.

Practically it means that the read out must have a very large dynamic range to accommodate these signal levels. Traditionally a digitizer with 14 bit resolution was chosen. For data taking this is adequate, but for $\gamma$-ray calibration with sources such as  $^{137}$Cs at 662 keV the S2 signal might be out of range. Reducing the overall sensitivity of the read out is not an option since low energy S1 from DM candidates would no longer be detectable.

The CAL group measured the S2 gain in their test to 287$_{-75}^{+97}$ Ph/e$^{-}$. At this gain S2 pulses are equivalent in amplitude to their dual phase detector. To reach this gain they had to increase the anode potential beyond the avalanche threshold. Thus, their gain includes a factor 14 from electron multiplication. If we remove this factor we expect a S2 gain of about 20 Ph/e$^{-}$. The lower gain does not mean a lower resolution since the statistic of the measurement is controlled by the number of drifting electrons which remains unaltered. In the data analysis most calculations involve both the S2 and S1 signals. In all these cases the weaker S1 signal dominates the error. 

Recently there is a heightened interest in high energy events, e.g. from neutrino-less Double Beta Decay of  $^{136}$Xe with a Q-value of 2458 keV. Since the natural abundance of  $^{136}$Xe is 8.9 \%, present and future LXe DM detectors contain a large amount of this isotope. Even at such high energies the two electrons will form a single-site event, and the length of the track is much smaller than the spatial resolution of the TPC read out. Due to the reduced S2 gain the signals can now all be within the limits of the electronics with no saturation.

\section{Conclusion and Outlook}
Dual Phase LXe TPCs are very powerful tools. In recent years they have tremendously advanced the search for Dark Matter. The Dual Phase approach is an ingenious way to achieve sensitivity to low energy charge measurements impossible even with the best Charge Sensitive Amplifiers. In future very large detectors, however, the Dual Phase technique will be difficult to implement.

Proportional scintillation around thin wires in liquid xenon was first observed 40 years ago. All the technological problems were solved in the meantime. It now can be used in Single Phase detectors observing both charge and light. The method offers several unique features which would be beneficial in future very massive LXe detectors. The method using proportional Scintillation in the LXe was again studied by two teams from the Columbia Astrophysics Lab and from Shanghai Jiao Tong University. Although there was general agreement some effects were reported which might restrict the resolution and sensitivity. We were able to fully understand and explain the discrepancies in the results. The negative effects can be avoided in very large LXe TPCs and are harmless. For the next generation of LXe WIMP detectors we expect therefore a substantially easier design, a better performance, and a higher sensitivity over a much larger energy range.

Using Proportional Scintillation in the liquid, it also will be possible to extend experiments to much higher energies like in the search for neutrino less Double Beta Decay of $^{136}$Xe. This could increase the physics reach of LXe DM detectors, since with a natural abundance of 8.9 \% a detector with 4-ton active mass contains 350 kg of $^{136}$Xe.

\section{Acknowledgment}
This project has been supported by a grant from the Ministry of Science and Technology of China (Grant NO.2016YFA0400301).

\end{document}